\documentclass[trackchanges,twocolumn]{aastex}

\let\tablenum\relax

\usepackage{natbib,epsfig,siunitx,url,graphicx,amsmath,footnote,float,xcolor,cases}
\usepackage{newtxtext,newtxmath}
\usepackage{siunitx}
\usepackage[T1]{fontenc}
\usepackage{tablefootnote}
\begin{document}

\submitted{Accepted for publication in MRNAS}

\title{Mathematical encoding within multi-resonant planetary systems as SETI beacons}

\author{Matthew S. Clement\altaffilmark{1}, Sean N. Raymond\altaffilmark{2}, Dimitri Veras\altaffilmark{3,4,5} \& David Kipping\altaffilmark{6}}

\altaffiltext{1}{Earth and Planets Laboratory, Carnegie Institution for Science, 5241 Broad Branch Road, NW, Washington, DC 20015, USA}
\altaffiltext{2}{Laboratoire d'Astrophysique de Bordeaux, Univ. Bordeaux, CNRS, B18N, all{\'e} Geoffroy Saint-Hilaire, 33615 Pessac, France}
\altaffiltext{3}{Centre for Exoplanets and Habitability, University of Warwick, Coventry CV4 7AL, UK }
\altaffiltext{4}{Centre for Space Domain Awareness, University of Warwick, Coventry CV4 7AL, UK}
\altaffiltext{5}{Department of Physics, University of Warwick, Coventry CV4 7AL, UK}
\altaffiltext{6}{Dept. of Astronomy, Columbia University, 550 W 120th Street, New York NY 10027}
\altaffiltext{*}{corresponding author email: mclement@carnegiescience.edu}

\begin{abstract}

How might an advanced alien civilization manipulate the orbits within a planetary system to  create a durable signpost that communicates its existence? While it is still debated whether such a purposeful advertisement would be prudent and wise, we propose that mean-motion resonances between neighboring planets -- with orbital periods that form integer ratios -- could in principle be used to encode simple sequences that one would not expect to form in nature. In this Letter we build four multi-resonant planetary systems and test their long-term orbital stability. The four systems each contain 6 or 7 planets and consist of: (i) consecutive integers from 1 to 6; (ii) prime numbers from 2 to 11; (iii) the Fibonacci sequence from 1 to 13; and (iv) the Lazy Caterer sequence from 1 to 16.  We built each system using N-body simulations with artificial migration forces. 
We evaluated the stability of each system over the full 10 Gyr integration of the Sun's main sequence phase.  We then tested the stability of these systems for an additional 10 Gyr, during and after post-main sequence evolution of the central stars (assumed to be Sun-like) to their final, white dwarf phase.  The only system that was destabilized was the consecutive integer sequence (system i). The other three sequences therefore represent potential SETI beacons.


\end{abstract}


\section{Background} \label{sec:intro}




Over the last few decades, the search for intelligent life within the cosmos has evolved from a primarily radio-centric enterprise (e.g. \citealt{cocconi:1959,drake:1961,tarter04}) to one that includes an increasingly broad range of possible technosignatures \citep{techno18}. The longevity of such signatures is directly proportional to their detectability \citep{balbi:2021} and thus those that could persist for Myr, or even Gyr, are of particularly keen interest. However, this raises questions about what kind of technosignature could feasibly produce a detectable signature for so long, especially when one recognises that most indications of human technology would likely be undetectable within a few million years \citep{schmidt:2019}.

Perhaps a sufficiently advanced civilization could engineer an active mechanical/electrical system able to reliably function over a Gyr or more, but an alternative approach is to consider a passive system inherently more robust to entropy's penchant for destruction. In the case of deliberate messaging, one proposed example is the construction of artificial transiting screens \citep{arnold:2005}. Although the longevity of such structures is unclear, and surely sensitive to the specific design and environment considered, the transmission energy is essentially provided by the star. By exploiting this astrophysical object, the power source is thus \textit{inherently} long lived. However, the screens themselves may not be; potentially suffering degradation from dust abrasion \citep{hoang:2017,muller:2019}, radiation damage, or orbital instability \citep{batygin:2008}. This raises the question - in the same way that the power source can be astrophysical, and thus long-lived, could one use other astrophysical objects in place of the screens to foster longevity?

To serve as a technosignature though, the astrophysical object(s) in question would need to somehow clearly encode a signature of non-natural origin. Since the objects themselves are by definition astrophysical, inspection of their individual nature would not be sufficient. Instead, a degree of non-naturalness would have to be achieved through their \textit{arrangement}. The orbital architecture of companion planets thus emerges as a possible solution. In particular, we propose that multi-resonant planetary systems could serve this role, since such configurations can be persistent over Gyr and can encode a series of integers. Each pair of neighboring planets forms a mean motion resonance with an orbital period ratio that is the fraction of two integers. A chain of resonances thus encodes a sequence; one that could potentially signal intelligence (e.g. a mathematical sequence; \citealt{sagan:1975}) and be arranged through local manipulation (e.g. see \citealt{korycansky:2001}).

Of course, one challenge is that chains of orbital resonances frequently form in nature. Jupiter's Galilean moons are the first-discovered example, and the Trappist-1 exoplanet system is the longest currently-known (7-planet) resonant chain~\citep{gillon17,luger17_res}. Resonant chains are naturally produced during planet formation as a consequence of orbital migration in gaseous protoplanetary disks~\citep{lee02,kley12}. However, simulations show that not all resonances are populated equally: first-order resonances (in which the integer sequence differs by 1, such as the 4:3, 3:2 or 2:1 resonances) are by far the most common outcomes~\citep[][]{terquem07,pierens08,izidoro17}. Chains of 3:2 or 2:1 resonances are thus expected to form naturally and examples have indeed been discovered~\citep[e.g., the K2-138 chain of 3:2 resonances and the HR8799 chain of 2:1 resonances;][]{christiansen18,marois10}. Extended chains encoding abstract mathematical sequences have not been discovered and thus their detection would be quite curious. Two basic questions emerge concerning such sequences: i) how often would they naturally form (the false-positive rate to a technosignature search), and, ii) how long-lived could we expect them to be?

In this work, we limit our scope to the second of these questions. After all, if such sequences are not long-lived, their spurious formation would be inconsequential anyway. Our approach is to first build four resonant chain planetary systems (Section 2) that each encode a specific integer sequence (see Table \ref{table:chains}). While we assume that such a signpost could in principle have been produced by a highly-advanced civilization, we expect its later evolution to be passive and governed by natural laws.  However, it is worth acknowledging that an alien species advanced enough to harvest the substantial quantity of energy necessary to assemble such a resonant signpost might also be capable of devising a long-lasting system that would prevent orbital instabilities by applying slight deviations to the planets' orbits.  Nevertheless, we proceed by testing the dynamical stability of each system during its central star's 10 Gyr main sequence lifetime (Section 3) and the next 10 Gyr of post-main sequence evolution (Section 4).  We discuss the interpretation of such signatures in Section 5.

\section{Building resonant chains}


Resonant planetary systems are naturally produced during gas-driven migration~\citep{lee02,batygin13,pichierri18}.  We leverage the standard approach of modifying the equations of motion to mimic gas disk-induced orbital migration by including forced migration ($\dot{a}$) and eccentricity damping ($\dot{e}$) in the {\tt Mercury} N-body integration package \citep{chambers99}. We made use of \textit{Mercury's} Bulirsch-Stoer algorithm (with a tolerance setting of \num{1.0e-15}), as the force on any given particle is a function of both their position and momenta. The time-step was set to 1$\%$ of the shortest orbital period in the simulation, and the central star's mass to 1.0 $M_{\odot}$.

To produce multi-resonant planetary systems, we initialized each planet from the inside-out on a circular, co-planar orbit with a semi-major axis just outside of the desired resonant semi-major axis (with a period ratio $\sim$1$\%$ larger than the desired final period ratio; see Table \ref{table:chains}).  For consistency, each of our resonant chains places the innermost planet at 5.0 au.  This selection ensures none of the planets will be engulfed during the AGB evolutionary phase. We then applied migration forces, integrating each system up to $t=$ 10 Myr. We verified that pairs of planets are indeed in the desired resonance by inspecting resonant angles of the form:
\begin{equation}
	\phi = p \lambda_{2} - q \lambda_{1} - r \varpi_{2} - s \varpi_{1},
	\label{eqn:res}
\end{equation}
\noindent where $\lambda_{i}$ and $\varpi_{i}$ are the mean longitudes and longitudes of perihelia of each planet (with planet 2 being the outer one).  The integers p and q define the resonance's period ratio, p:q, and the remaining integers satisfy:
\begin{equation}
	p=q+r+s.
\end{equation}
In resonance the values of $\phi$ librate about a specific value as the consequence of the fact that resonant entrapment shields objects from close approaches by coupling their orbital and precession frequencies (see Fig.~\ref{fig:10gyr}).

We iterated through migration simulations until the desired resonance was established and remained intact throughout the initial 10 Myr simulation. Several migration simulation attempts are often required to determine input parameters that prevent the distant planets from being captured in powerful first-order resonances (e.g.: 3:2 instead of 11:7), however we do not see this as a major obstacle for an advanced civilization.  For additional details on our numerical pipeline for generating unique resonant chains, we direct the reader to \citet{clement21_instb}.

\begin{table}
    \centering
    \begin{tabular}{c|c|c|c}
    \hline
    Name & $N_{pln}$ & Resonant Chain & $M_{pln}$ ($M_{\oplus}$) \\
    \hline
    Prime 1 & 6 & 1:2:3:5:7:11 & 20,20,20,20,2,2 \\
    Prime 2 & 6 & 1:2:3:5:7:11 & 2,2,2,2,2,2 \\
    Fibonacci 1 & 7 & 1:1:2:3:5:8:13 & 20,20,20,20,20,2,2 \\
    Fibonacci 2 & 7 & 1:1:2:3:5:8:13 & 2,2,2,2,2,2 \\
    Lazy Caterers & 6 & 1:2:4:7:11:16 & 20,20,20,20,2,2 \\
    Integer & 6 & 1:2:3:4:5:6 & 20,20,20,20,2,2 \\ 
    \hline
    \end{tabular}
    \caption{Summary of resonant chains tested in our investigation.  The columns are as follows: (1) The simulation name, (2) the total number of planets in the chain, (3) the orbital period ratios of each consecutive pair of planets from inside out (i.e: $P_{1}$:$P_{2}$:$P_{3}$, etc), and (4) the mass of each planet starting with the innermost body in earth units.}
    \label{table:chains}
\end{table}


\section{Main Sequence stability}

We tested the dynamical stability of each of the resonant chains (see Table~\ref{table:chains}) over the full 10 Gyr duration of their (Sun-like) central star's main sequence lifetime.  We used the Hybrid integrator in the {\tt Mercury} integration package~\citep{chambers99}, again using a timestep that was 1\% of the innermost planet's orbital period.

\begin{figure*}
    \centering
    \includegraphics[width=.45\textwidth]{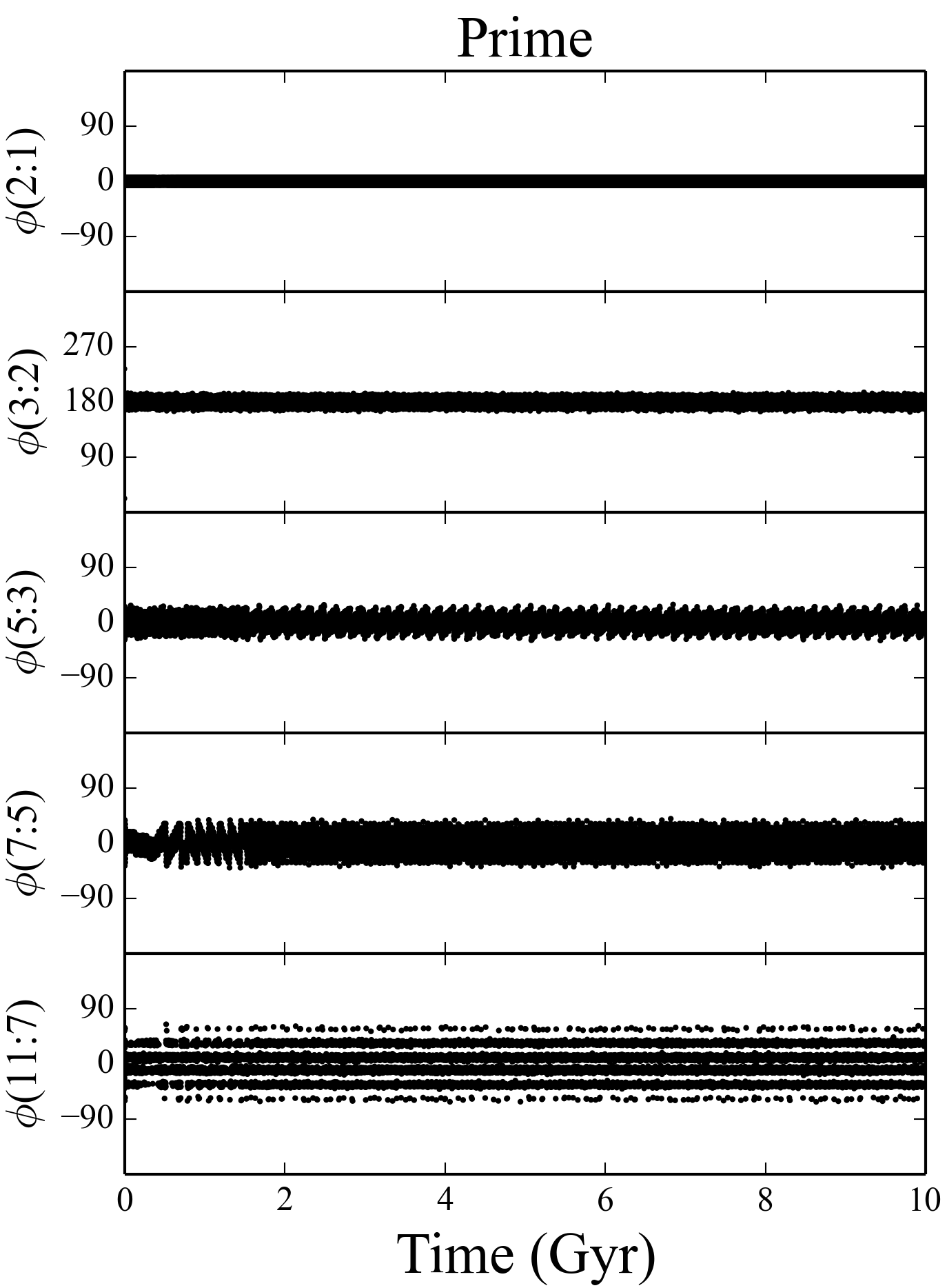}
    \includegraphics[width=.45\textwidth]{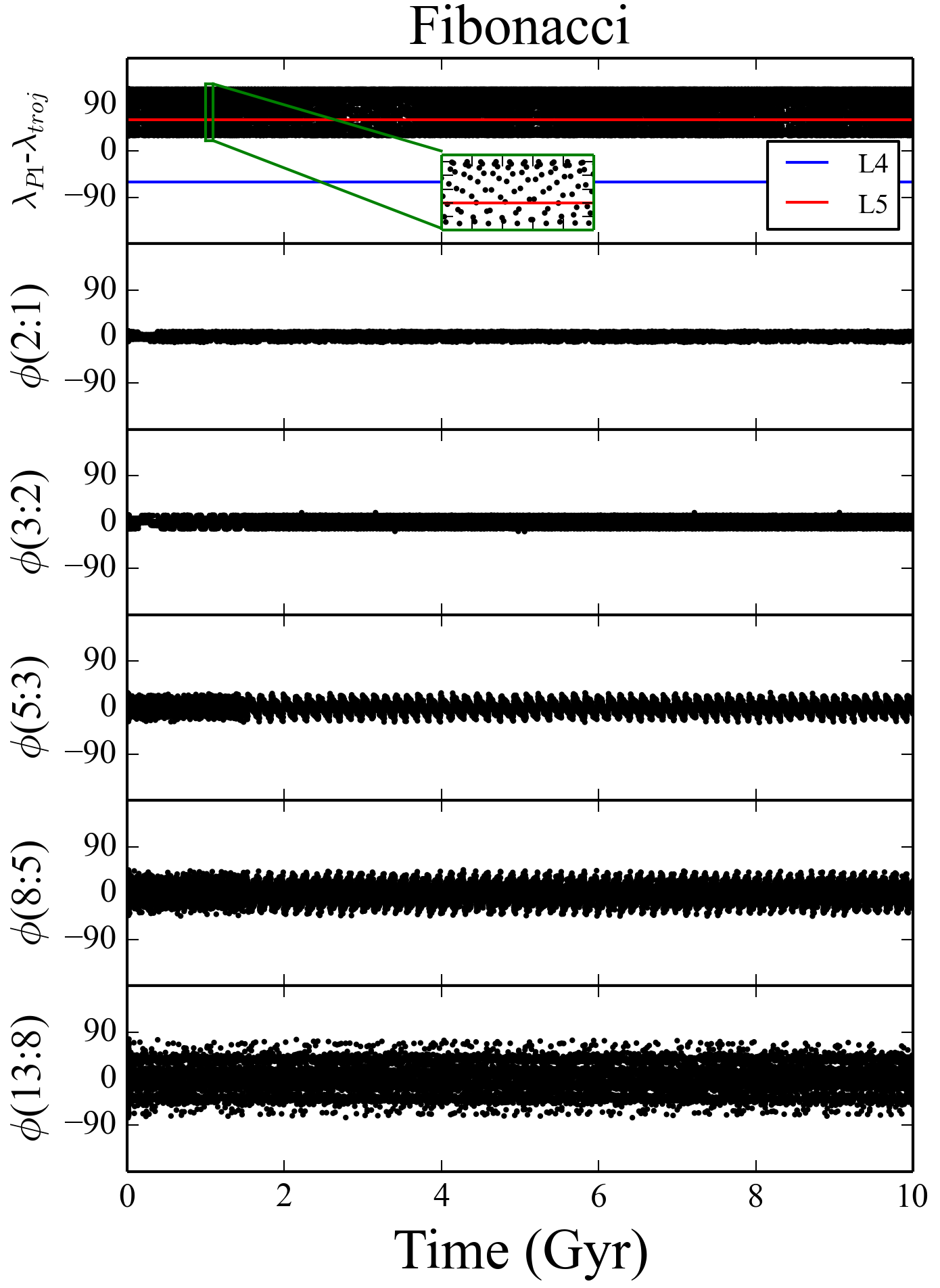}
    \qquad
    \includegraphics[width=.45\textwidth]{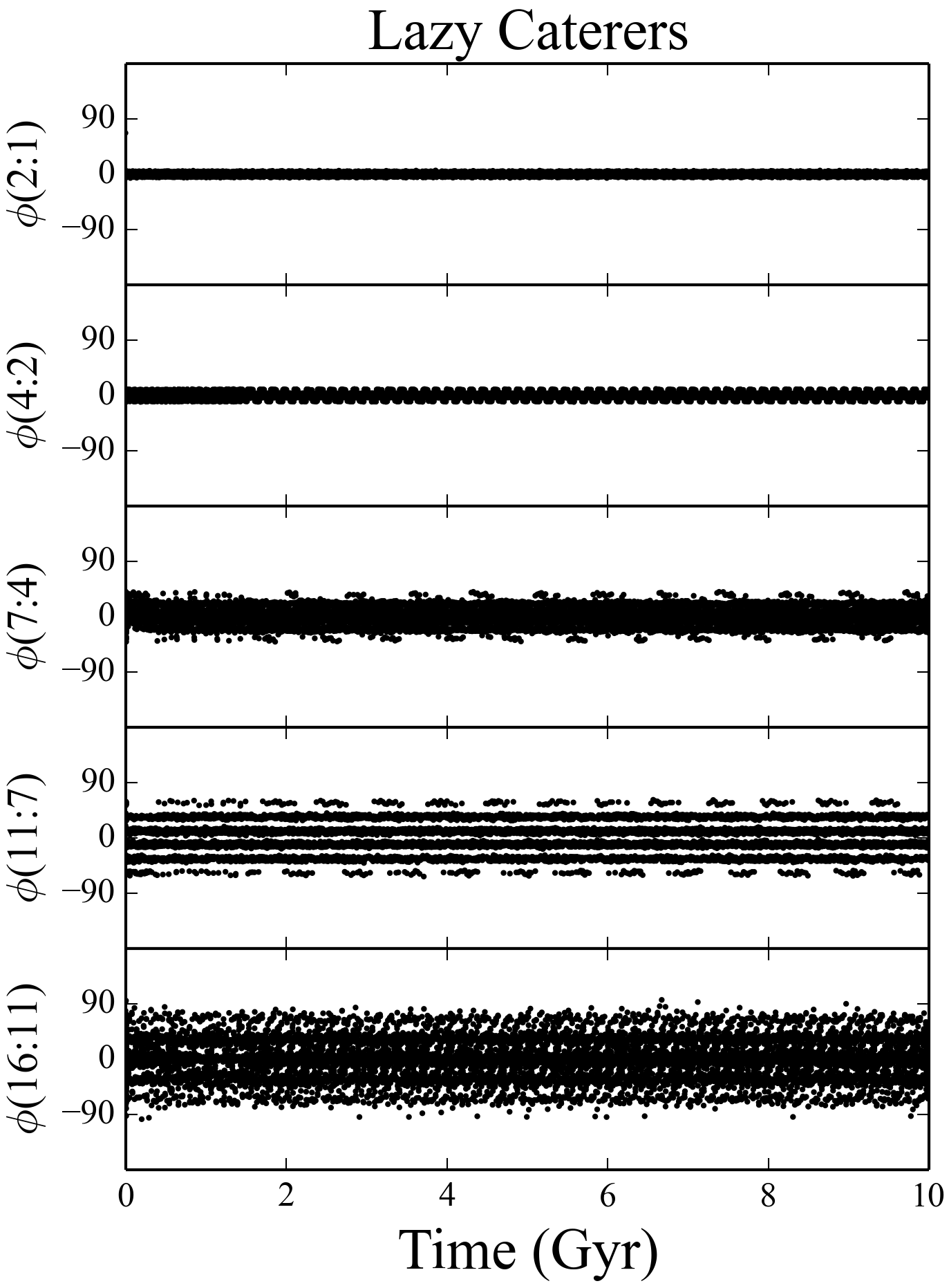}
    \includegraphics[width=.45\textwidth]{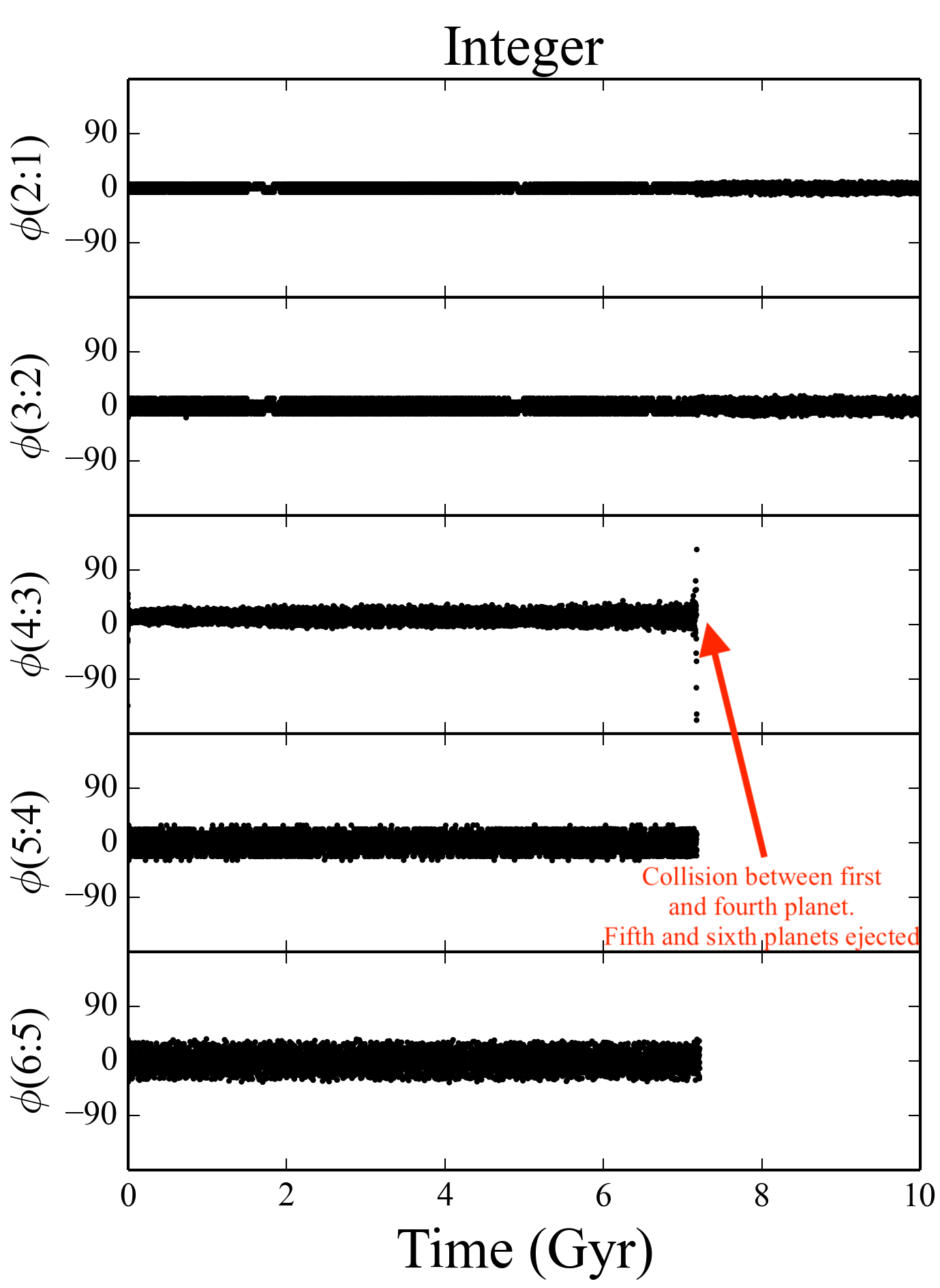} 
    \caption{Resonant angle evolution in our 10 Gyr simulations of our Prime 2 (upper left panel) Fibonacci 2 (upper right panel), Lazy Caterers (lower left panel) and Integer (lower right panel) resonant chains (table \ref{table:chains}).  Each successive subplot depicts the dominant resonant angle (equation \ref{eqn:res}) for each respective pair of planets, starting with the innermost object.  The top subplot in the upper right panel (Fibonacci chain) and the accompanying insert depict the position of the Trojan planet with respect to the inner planet (i.e.: in the rotating reference frame: $\lambda_{1}-\lambda_{2}$), along with the positions of the L4 and L5 Lagrange points.}
    \label{fig:10gyr}
\end{figure*}

Figure \ref{fig:10gyr} plots the 10 Gyr evolution of the relevant resonant angles (equation \ref{eqn:res}) of our four resonant chains (see Table \ref{table:chains}).  The only chain to destabilize was the more compact chain of consecutive integers.  In this case, the outer 4 planets spontaneously dislodged from resonance around 7.2 Gyr into the simulation.  Once the resonant phase protection mechanism terminated, all six planets began to interact strongly; thus exciting the eccentricity of each body to $\sim$0.20 within the next 2 Myr.  After this occurred, the first and fourth planets collided and the fifth and six planets were ejected after scattering off of the third planet.  

Our five other chains each remained stable for 10 Gyr, with resonances maintained. Of course, while this demonstrates the general integrity of these configurations, it does not prove that they would evolve regularly in the presence of additional planets and small bodies.  Indeed, perturbations from small bodies and additional non-resonant (or resonant) planets are known to drive instabilities in systems of otherwise stable planets both in the solar system \citep{nesvorny12} and elsewhere \citep[e.g.:][]{raymond10,izidoro17}.

\section{Post-Main Sequence evolution}

To propagate our simulations beyond the main sequence, we applied the $N$-body code described in \cite{musetal2018}. This code incorporates stellar evolution output from the {\tt SSE} code \citep{huretal2000} into the framework of the RADAU integrator from the {\tt Mercury} software package \citep{chambers99}. Our code is the same one used to determine the post-main-sequence breaking of the four-planet resonant chain in the HR 8799 system \citep{verhin2021}.

The physical and chemical properties of the parent star determine its evolutionary profile. Because here we are just seeking to demonstrate a proof-of-concept, we used a single set of fiducial parameters. In addition to adopting a main-sequence stellar mass of $1.0M_{\odot}$, we also assumed Solar metallicity, a Reimers mass loss coefficient of 0.5 \citep{reimers1975}, and included the superwind phase along the asymptotic giant branch \citep{vaswoo1993}. 

We integrated the resulting stellar evolutionary mass profile with each of the systems in Table \ref{table:chains} in three separate simulations: one with a RADAU tolerance of $10^{-10}$, another with a tolerance of $10^{-11}$ and a third with $10^{-12}$. These tolerances determine the variable timesteps by which the planets are propagated, and hence our collection of three simulations per architectural setup provides some small statistical measure of the fate of these systems. 

We integrated the systems from the end of the main-sequence phase for 10 Gyr. This timespan includes the approximately 1.4 Gyr duration traversing the giant branch phases, about 0.2 Myr of which covers the asymptotic giant branch phase. In all cases, the planets begin sufficiently far away from the star such that there is no danger of engulfment into the expanding giant star unless one the planets is scattered inward due to an instability.

The majority of architectures from Table \ref{table:chains} remained stable for the duration of our simulations. In fact, gravitational instability appeared in only two of the chains: (i) the Fibonacci 1 chain, and (ii) the Integer chain. For (i), all three simulations remained stable throughout the giant branch phases, and then became unstable approximately 2.0, 2.4 and 5.2 Gyr into the white dwarf phase. For (ii), the instability timescale and trigger were qualitatively different: here the end of the asymptotic giant branch phase triggered instabilities quickly, just 87, 148 and 464 Myr after the star became a white dwarf.

Unlike the instability depicted in Fig. \ref{fig:10gyr}, each post-main Sequence instability resulted in a destruction of the entire resonant chain.  We also checked for libration of the relevant resonant angles in our stable evolution and found that the first order resonances (i.e.: 2:1, 3:2, 4:3, etc.) remained intact for the entire 10 Gyr calculation.  While the libration of the higher order resonant angles was lost during the giant branch phases in the majority of our simulations, in all stable cases the planets' period ratios remained within $\sim$0.05$\%$ of their initial values for the duration of the integration. We note that, in any case, it would not be possible to detect the libration of a distant pair of planets in a higher order resonance with current technology. 

\begin{figure*}
    \centering
    \includegraphics[width=.45\textwidth]{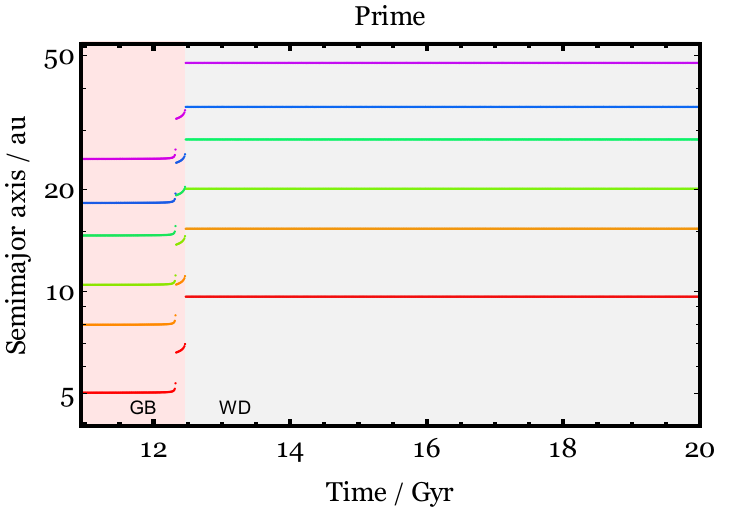}
    \includegraphics[width=.45\textwidth]{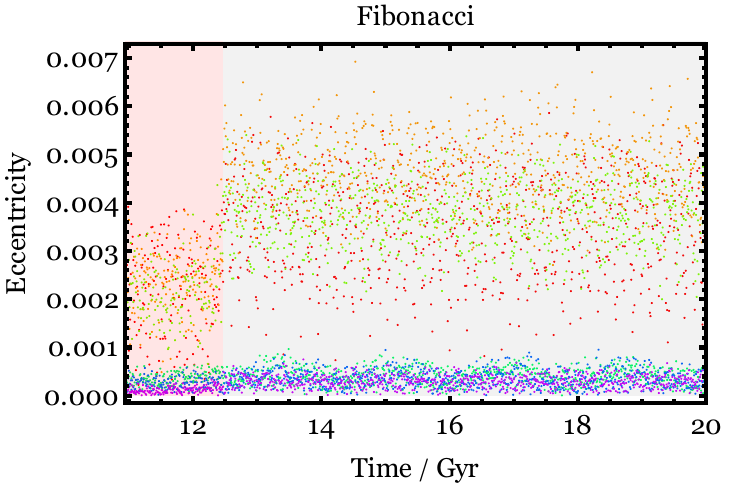}
    \qquad
    \centerline{}
    \includegraphics[width=.45\textwidth]{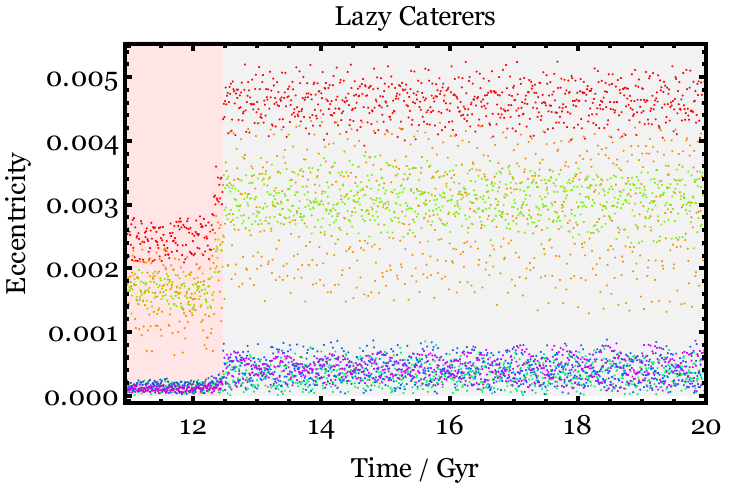}
    \includegraphics[width=.45\textwidth]{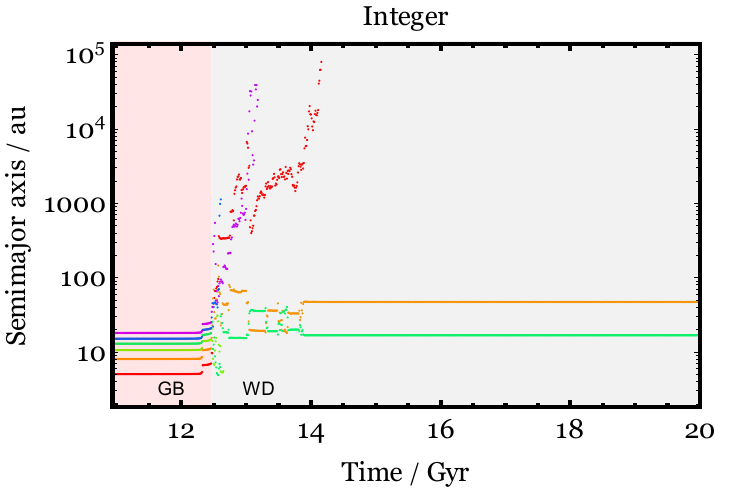} 
    \caption{Evolution during the giant branch (GB) and white dwarf (WD) phases of the same systems from Fig. \ref{fig:10gyr}, but starting at the end of the main sequence. The semi-major axis evolution is shown in the Prime 2 (upper left panel) and Integer (lower right panel) systems, and the eccentricity evolution is shown in the Fibonacci 2 (upper right panel) and Lazy Caterers (lower left panel) systems. The semi-major axis ratio between each pair of planets remains nearly fixed during stellar mass loss, and the corresponding eccentricity increases are on the order of $10^{-3}$. Only the Integer sequence becomes unstable, and just only about 100 Myr after the GB-WD transition.}
    \label{fig:10gyrPMS}
\end{figure*}

Schematically, some of the evolutions are shown in Fig. \ref{fig:10gyrPMS}. The figure shows how the semi-major axis ratio of each planet pair remains nearly fixed during stellar mass loss, and the orbital eccentricity increases are on the order of $10^{-3}$. These variations are natural and reflect the true nature of non-adiabatic stellar mass loss on orbiting objects \citep{veretal2011}, although the primary trigger for instability on these spatial scales probably represents the change in stability boundaries due to the altered gravitational potential \citep{debsig2002}.

\section{Discussion}

It is worth considering how confident we can be that our chosen resonant chains unequivocally signify the presence of an intelligence. As discussed above, chains of orbital resonances -- especially first-order resonances like the 2:1 and 3:2 -- are a natural outcome of both the planet- \citep[e.g.:][]{masset01} and satellite \citep[e.g.:][]{madeira21} formation processes. Notable examples include the three inner Galilean moons, the K2-138 system of close-in exoplanets~\citep{christiansen18}, and the HR 8799 system of wide-orbit gas giants~\citep{marois10}.  It is also likely that the Solar System's giant planets were in a primordial resonant chain when they emerged from the gaseous disk~\citep{morby07,nesvorny12,clement21_instb}.  The only examples of higher order resonances in the solar system are in small body and satellite reservoirs.  For instance, Neptune's moons Naiad and Thalassa are in a 73:69 resonance \citep{brozovic20}, and trans-Neptunian objects have been detected in resonances as distant as the 9:1 \citep{volk18} and 11:1 \citep{clement21_p9}.  High-order resonances are difficult to confirm in exoplanet systems; they are expected as an outcome of dynamical instabilities but only at a very low rate and in systems with large orbital eccentricities~\citep{raymond08}.

An effective SETI beacon would encode a mathematically-meaningful sequence that includes multiple resonances more exotic than the common 2:1 and 3:2.  Ideally, such a beacon would include at least one extremely high-order resonance (such as the 11:7 or 13:8 in our Prime and Fibonacci chains) and avoid sequences that grow exponentially (thus making the distant bodies extremely difficult to detect).  In this manner, a sequence like our Fibonacci chain that includes a pair of co-orbital (Trojan) planets would be extremely effective as a signpost of an alien civilization, as such configurations are not expected to be likely end states of planet formation.\footnote{The Fibonacci sequence was indeed the one transmitted by intelligent aliens in {\em Contact}~\citep{sagan85}.} Given that the exact period ratios of planets can deviate from perfect integer ratios, it is important that a SETI beacon system be characterized by librating resonant angles (as in Fig.~\ref{fig:10gyr}). 

Might any currently-known resonant chains qualify as possible SETI beacons? The best current candidate is the TRAPPIST-1 system, whose 7 planets form an 8:5;5:3;3:2;3:2;4:3;3:2 resonant chain~\citep{gillon17,luger17_res}.  If we require the innermost planet's orbital period to be an integer, this constitutes the following sequence from the innermost planet-outward: 15,24,40,60,90,120,180. While this sequence is of no obvious mathematical significance, a simpler sequence can be constructed from the outermost planet-inward: 2,3,4,6,9,15,24. This represents consecutive values in the sequence representing the number of compositions of odd numbers into primes of the form $x^2+y^2$. While this is not an integer sequence\footnote{See \url{https://oeis.org/A287148}} of note, we cannot rule out the possibility of a SETI beacon. However, it has been proposed that each pair of planets was initially captured into first-order (3:2 or 4:3) resonances~\citep{papaloizou18}.  It is also worth noting that this sequence could be extended in either direction.  While the existence of an additional, undetected interior planet in the system is unlikely, an outer sub-giant planet in a 2:1 resonance with  TRAPPIST-1h cannot be ruled out by current available radial velocity data \citep[e.g.:][]{boss17}.  However, such a planet would have to be less than a few Earth masses so as to not overly excite the inner planets' eccentricities \citep{dencs19} beyond the observed values \citep{quarles17,grimm18}.

While the energy required to modify the orbits of a collection of planetary-mass objects by several astronomical units is substantial, a civilization capable of exploiting the complete energy output of its host-star \citep[e.g., a Kardashev Type II civilization:][]{kardashev64} could acquire the necessary energy in a reasonable amount of time.  Indeed, the orbital energy needed to modulate the semi-major axes of each of the six planets in our Prime 2 sequence by $\sim$10$\%$ is around \num{2.8e32} $J$.  Even if an  advanced civilization around a 1.0 $M_{\odot}$ star were only 10$\%$ efficient at capturing solar energy (for instance, by utilizing a Dyson sphere), it would only require 2.3 years to build up the necessary energy.  While this timeline would be lengthened by around two orders of magnitude if the civilization happened to inhabit an early M-Dwarf system, it is also possible such a species would employ more passive means of orbital modification.  Indeed, it is feasible to devise orbital trajectories for smaller, $\sim$asteroid-mass objects that slowly transfer energy between two planets in a system over longer periods of time, thus modifying the orbits of both planets in opposite directions with relatively minimal energy input to the 3-body system \citep{korycansky:2001}.

We conclude by emphasizing the remarkable scientific value of resonant chain planetary systems. Multi-resonant systems induce larger-amplitude transit-timing variations than their non-resonant counterparts~\citep{agol05}, enabling precise mass determination~\citep[][]{agol21} and constraining the planets' bulk compositions~\citep{dorn15} and bombardment histories~\citep{raymond21}. As SETI beacons, resonant chains encode simple integer sequences that can be deciphered by stars able to see those systems edge-on~\citep[see][]{kaltenegger21}. Finally, as we have shown, a carefully-constructed resonant beacon can in principle remain dynamically stable for longer than the current age of the Universe.

\section{Acknowledgements}

The authors would like to thank Alexander Mustill for providing a thoughtful and comprehensive review that greatly improved the quality of the manuscript.  The authors also acknowledge Omar E. Pol, associate editor of the Online Encyclopedia of Integer Sequences, who first recognized the connection between TRAPPIST-1 and the A287148 series.   DV gratefully acknowledges the support of the STFC via an Ernest Rutherford Fellowship (grant ST/P003850/1).  SNR is grateful to the CNRS's {\em Programme Nationale de Plantologie (PNP)}. DK thanks supporters to the Cool Worlds Lab, Mark Sloan, Laura Sanborn, Douglas Daughaday, Andrew Jones, Marc Lijoi, Elena West, Tristan Zajonc, Chuck Wolfred, Lasse Skov, Alex de Vaal, Mark Elliott, Methven Forbes, Stephen Lee, Zachary Danielson, Chad Souter, Marcus Gillette, Tina Jeffcoat, Jason Rockett, Scott Hannum, Tom Donkin, Andrew Schoen, Jacob Black, Reza Ramezankhani, Steven Marks, Philip Masterson, Gary Canterbury \& Nicholas Gebben. This paper was inspired by a blog post that we wrote entitled `Cosmic Time Capsules' (\url{https://planetplanet.net/2021/10/25/cosmic-time-capsule/}).

\section{Data Availability}

The simulation inputs and results discussed in this paper
are available upon reasonable request to the corresponding
author.

\newcommand{\sci}{$Science$ }
\newcommand{\psj}{$PSJ$ }
\bibliography{bib}{}
\bibliographystyle{apj}

\end{document}